# An AI-guided mechanotyping instrument for fully automated oocyte quality assessment

*Yining Guo#，Wenshuo Zhao#，Xueying Sun#，Jing Huang，Xi Chen，Xinyu Lu, Yuan Liu，Haifeng Xu\**

Y. Guo, W. Zhao, X. Sun, J. Huang, X. Chen, X. Lu, Y. Liu, H. Xu

Shenzhen Institutes of Advanced Technology, Chinese Academy of Sciences, Shenzhen 518055, China

E-mail: hf.xu@siat.ac.cn

W. Zhao

College of Mechanical Engineering, Hubei University of Technology, Hubei 430068, China

X. Chen

College of Electronic and Optical Engineering & College of Flexible Electronics (Future Technology), Nanjing University of Posts and Telecommunications, Nanjing 210023, China

Funding:
National Natural Science Foundation of China 32422046, National Natural Science Foundation of China 52303167, National Natural Science Foundation of China 52203152. Shenzhen Science and Technology Program JSGGKQTD20221101115654021 and RCBS20221008093222008.

**Abstract**

The mechanical properties of oocytes are regarded as important indicators of their developmental potential. During fertilization, deviations from the normal mechanical range can hinder sperm penetration, ultimately reducing fertilization efficiency and compromising embryo quality. However, current methods for measuring oocyte mechanics often suffer from serious cellular damage, low automation levels, and large measurement errors. To address these limitations, we developed an AI-guided μN-scale mechanical measurement system for safe and automated oocyte quality assessment. The system integrates voice interaction with automated experimental workflows to control a magnetically actuated microgripper, which applies defined loading forces to induce micron-scale compressive deformation of the oocyte. Combined with AI-assisted object detection and image segmentation algorithms, the system captures cellular deformation in real time, enabling precise calculation of the oocyte's compressive modulus. This measurement system enables automated, quantitative, and non-destructive evaluation of oocyte mechanical properties, providing an effective approach for oocyte quality screening in in vitro fertilization (IVF) and other assisted reproductive technologies (ART).

Y.G., W.Z. and X.S. contributed equally to this work.

**1. Introduction**

The quality and maturity of oocytes directly influence fertilization outcomes and subsequent embryo development, ultimately affecting clinical pregnancy rates[1–5]. Increasing evidence in recent years has demonstrated that the mechanical properties of oocytes are closely linked to their developmental potential.[4,6,7] Mechanical parameters such as cytoplasmic flow velocity,[8] cortical tension,[9] and compressive modulus[10,11] have been shown to correlate strongly with oocyte maturation status, fertilization competence, and embryo developmental trajectory. Among these parameters, the compressive modulus is considered a key parameter that reflects the structural integrity and biochemical composition of the oocyte.[4] Traditional mechanical measurement methods for oocytes mainly include atomic force microscopy (AFM) and micropipette aspiration (MPA), as reported by Gavara et al.,[12] Yanez et al.,[6] and other groups. AFM measures the elasticity of the cell surface or cortex through local probe contact, and its results are highly dependent on the geometry of the probe, contact model, and indentation depth.[12,13] Measurement errors could occur when the indentation is too deep or when there is interference from rigid substrates.[14,15] Moreover, since oocytes are non-adherent cells, AFM measurements typically require prior physical fixation,[13,16] which further increases the difficulty of experimental procedures. MPA, on the other hand, uses negative pressure to aspirate the cell membrane or a portion of the cytoplasm into a micropipette, which poses a high risk of membrane damage to the oocyte.[6,17] In addition, both AFM and MPA are strongly operator-dependent, with a limited level of automation, which restricts their application in laboratory workflows and large-scale clinical screening. More importantly, the mechanical parameters measured by these two methods primarily reflect local structural characteristics, making it difficult to directly characterize the overall mechanical behavior of the oocyte.[18–20]

In recent years, a number of studies have sought to develop measurement approaches that more accurately reflect the global mechanical properties. Lange et al. developed a high-throughput microfluidic constriction assay in which suspended cells are driven through an array of narrow constrictions. However, cells in this configuration must undergo large, flow-induced deformations to pass a narrow channel that are significantly smaller than their resting diameter, a loading regime that is likely incompatible with fragile oocytes. Barbier et al.[4] also used a constriction-based microfluidic device to drive oocytes through microchannels with mild deformation, evaluating cell deformability and developmental potential through deformation dynamics of oocytes. However, in such microfluidic constriction channels, cells are subjected to axial stretching, sidewall shear, and contact friction, leading to an uncontrollable and non-uniform stress state.[21–23] Therefore, there is a strong demand for a non-destructive, automated method that can quantify the whole-cell compressive modulus of individual oocytes under defined loading for reliable oocyte mechanical phenotyping and quality assessment.[24]

Here, we report an AI-guided automated µN-scale mechanical measurement system to quantify the compressive modulus of oocytes at the whole-cell level for quality assessment. **(Figure 1)** The system exhibits fully automated operation: it can scan and identify oocytes in microscopic images, precisely navigate and position target oocytes within the working space, and apply programmable µN-scale loading force to oocytes through magnetic actuation. By combining a real-time AI-guided object detection and segmentation algorithm, the system continuously monitors oocyte deformation and synchronously records the applied force to calculate compressive modulus. Moreover, the magnetically actuated microgripper confines deformation within a non-destructive regime, allowing whole-cell mechanical assessment without compromising oocyte integrity. Using mouse oocytes as a model, we demonstrate that the system can reliably identify a normal oocyte compressive modulus range of 420–800 kPa. Overall, this measurement instrument provides an integrated, automated, and non-destructive approach for quantifying oocyte mechanical properties, with strong potential in the application of oocyte quality assessment within assisted reproductive technologies.

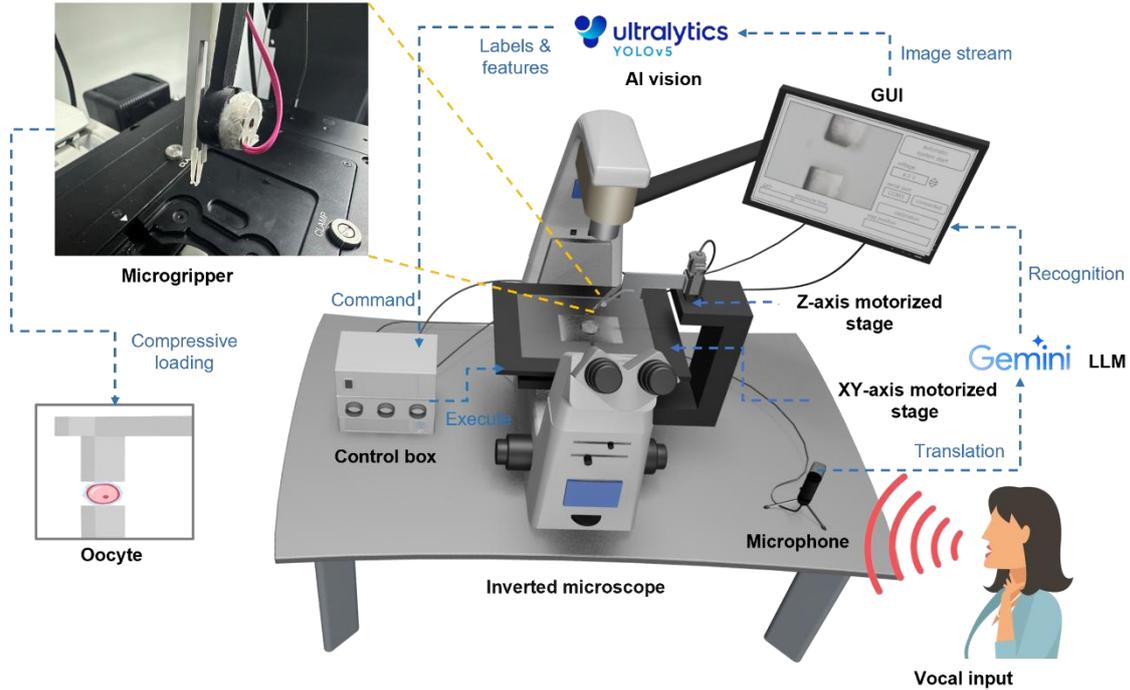

**Figure 1. Schematic diagram of the AI-guided automated µN-scale mechanical measurement system.** The system consists of a microgripper setup, a control box, an inverted microscope, a XY-axis motorized stage, a Z-axis motorized stage, a graphical user interface (GUI), a voice interaction module, and an AI-based computer vision algorithm.

## 2. Results and Discussion

### 2.1 Instrument Setup and Principles of Mechanical Measurement

The measurement instrument contains a microscopic imaging module, a magnetically actuated microgripper setup, an AI-based computer vision algorithm, a graphical user interface (GUI), a voice interaction module, a control box, and actuators for automated and safe oocyte compressive modulus measurement (**Figure 2a**). The microscopic imaging module provides stable optical conditions for microscale observation. An industrial camera is mounted on the microscope to capture real-time microscopic images. The magnetically actuated microgripper comprises a rigid arm, a flexible arm, and an electromagnetic actuation unit (Figure 2c). The rigid arm and the flexible arm of the microgripper possess gripper tips that are geometrically aligned in opposition to each other. Driven by the electromagnetic unit, the flexible arm undergoes controlled deflection, resulting in the defined loading force to the oocyte during mechanical measurement. The electromagnetic actuation unit, which consists of a cylindrical magnet fixed to the flexible arm and an electromagnet, controls the opening and closing of the gripper (Figure 2b). To achieve automated mechanical measurement of the oocyte compressive modulus, we integrate an AI-based computer vision algorithm, which is built on the Ultralytics

YOLOv5 framework[25]. The algorithm achieves high-precision oocyte recognition in microscopic images and automatically generates labels, position information, and segmentation features for the oocyte. The instrument supports both a graphical user interface (GUI) and vocal input for user interaction. Vocal input is converted into text and processed by the large language model (LLM), which identifies the valid phrase segments related to experimental commands within the voice input. Once the corresponding keywords are detected, the system automatically executes the fixed experimental procedure, enabling program initiation and oocyte target selection. All actuation commands in the workflow and data streams are ultimately routed to the control box, which is responsible for executing commands from control programs into physical operations through a series of high-precision actuators. The XY-axis motorized stage can be precisely controlled via a program to translate the oocyte sample, ensuring that individual oocytes are accurately aligned within the microscopic view and properly located. Installed on the top of the XY-axis motorized stage is the Z-axis motorized stage with the magnetically actuated microgripper setup mounted on it. This configuration provides the microgripper with a degree of freedom along the Z-axis, allowing it to remain retracted during sample positioning and to engage with the oocyte only when performing mechanical measurement operations, thereby minimizing unintended contact and disturbance to the oocyte.

In our system, the oocyte compressive modulus is derived from the deformation response under a well-defined loading force generated by the microgripper. In order to apply the loading force to the entire oocyte, gripper tips with a size of 400 μm were used, which is significantly larger than the oocyte diameter (80–100 μm). This ensures uniform compressive deformation of the cell, making it more suitable for non-destructive mechanical measurement at a whole-cell level. As shown in Figure 2d, the flexible arms of the microgripper deform under different externally applied magnetic fields, and the maximum opening distance $X$ between two microgripper tips can exceed 350 μm, covering the size range of the vast majority of mammalian oocytes and some small invertebrate oocytes.[26] During mechanical measurement, the automated control program gradually decreases the drive current of the electromagnet, leading to a reduction in the magnetic dipole force $F_{MAG}$ exerted on the cylindrical magnet mounted on the flexible arm. Consequently, the arm moves inward and applies a restoring force $F_R$ to the oocyte, and the force experienced by the oocyte $F_{cell}$ can be expressed as:

$$F_R + F_{MAG} + F_{cell} = 0 \tag{1}$$

To determine $F_{cell}$ quantitatively, we make use of the equilibrium relationships established during the calibration of the microgripper. In the absence of both the oocyte and calibration sensor, the static equilibrium between the restoring force of the flexible arm and the magnetic attraction between the electromagnet and the cylindrical magnet is given by

$$F_R + F_{MAG} = 0 \tag{2}$$

When the calibration sensor is introduced into the loading path during calibration, the equilibrium of the system becomes

$$F_R + F_{MAG} + F_S = 0 \tag{3}$$

where $F_S$ is the force measured by the calibration sensor.

Comparing Equations 1 and 3 under identical drive current and opening distance between microgripper tips shows that the loading state of the oocyte during mechanical measurement is mechanically equivalent to that of the calibration sensor. Therefore, the force experienced by the oocyte can be directly obtained from the calibrated force,

$$F_S = F_{cell} \tag{4}$$

In practice, the experimentally determined relationship between $F_S$, the drive current, and the gripper-tip displacement is used to convert the measured gripper kinematics into the effective oocyte loading force in all subsequent oocyte compression experiments.

**2.2 Instrument Calibration**

Prior to the mechanical measurement experiment, the magnetically actuated microgripper was calibrated to determine the relationship between the opening distance $X$ between microgripper tips, drive current, and the corresponding loading force. To verify the repeatability of the instrument under a fixed drive current of 18.13 mA, we manually measured the opening distance of the microgripper tips 50 times under the microscope (Figure 2e). The resulting measurements exhibit a high degree of consistency, with only small variations in the opening distance across the 50 trials (154.88μm ± 3μm). Figure 2f demonstrates the calibration setup based on a μN force sensor. To avoid interference during calibration, only the flexible arm of the microgripper is retained. To ensure measurement accuracy, the μN force sensor is employed. The flexible arm of the microgripper and the μN force sensor are arranged in a horizontally opposed configuration, allowing direct measurement of the loading force generated by the gripper tip. This configuration helps establish a force-displacement conversion relationship.

At the beginning of each calibration trial, a preset current $I$ is applied to the electromagnet, attracting the flexible arm upward and bending it to a certain position. Each initial current corresponds to a defined microgripper opening distance $X$. The microgripper tip is then brought into gentle contact with the µN force sensor, and the system is finely adjusted to ensure that the sensor reading is zero. With all components of the microgripper setup kept strictly fixed in space, the electromagnet drive current is then gradually decreased linearly. The reduction in current weakens the magnetic dipole force, allowing the flexible arm to rebound downward due to its elastic restoring force, thereby exerting an increasingly large force on the µN force sensor. This force is continuously measured and recorded by the sensor. We synchronously record a series of discrete current values and the resulting sensor readings, corresponding to different initial opening distances between microgripper tips. As shown in Figure 2g, for different initial microgripper opening distances ($X_1$ -$X_{12}$: 0-150µm), the force recorded by the µN sensor increases approximately linearly as the drive current decreases. However, small but systematic differences are observed between the individual curves: at the same drive current, larger initial opening distances correspond to higher measured forces. These results allow us to construct a set of calibration curves parameterized by the opening distance between microgripper tips, which relate the drive current to the measured force.

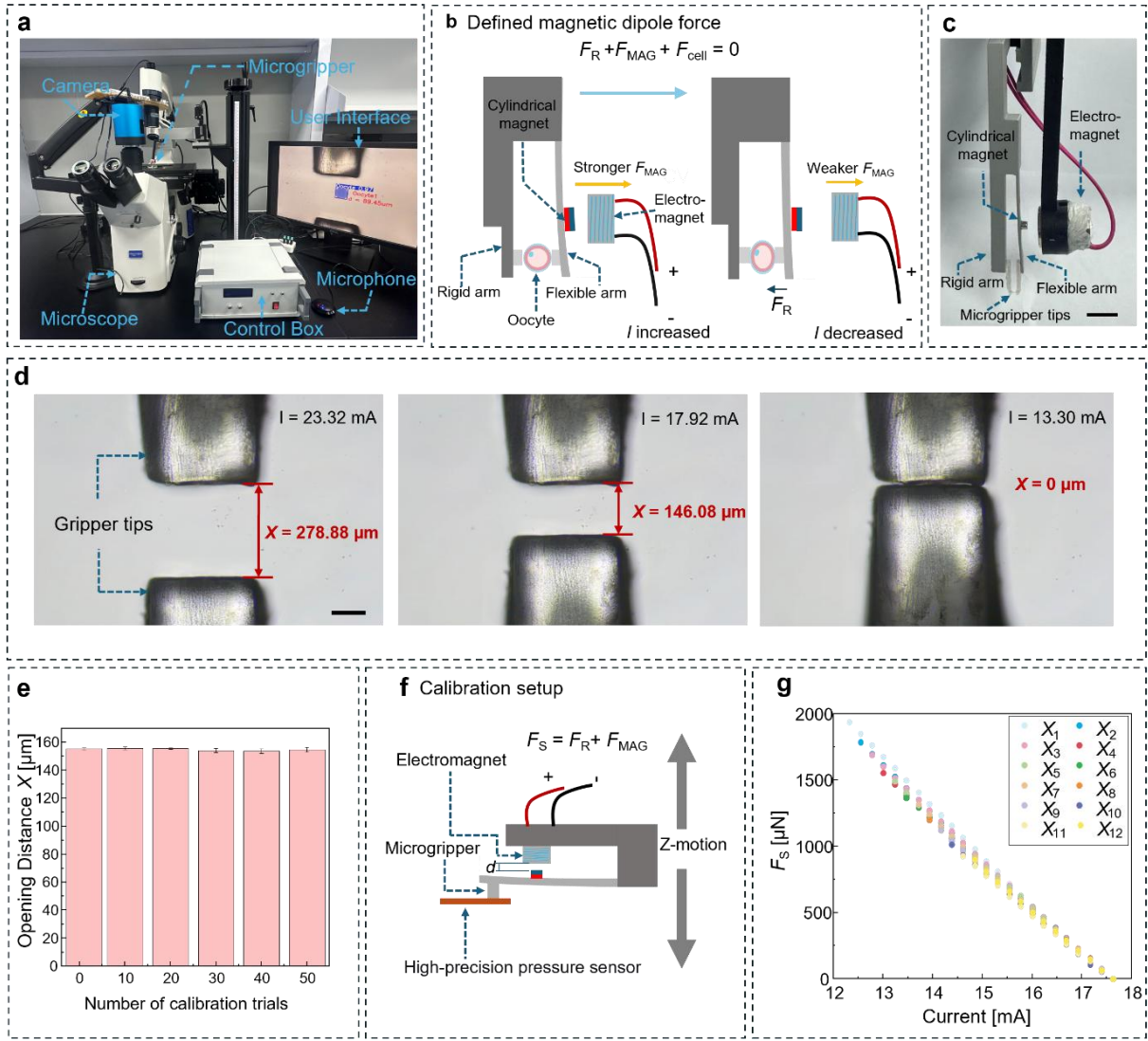

**Figure 2. Instrument setup and calibration.** a) Picture of the AI-guided μN-scale mechanical measurement instrument. b) Schematic of the microgripper setup and oocyte compressive deformation process. c) Picture of the microgripper setup (scale bar = 10 mm). d) Opening distance between microgripper tips under three different drive currents (scale bar = 100 μm). e) Repeatability test of opening distance under fixed drive current $I$ = 18.13 mA. f) Schematic of instrument calibration setup. g) Calibrated force-drive current relationship at different initial opening distances.

## 2.3 Oocyte Mechanical Measurement and Quality Assessment

To demonstrate the feasibility and versatility of the measurement system, mouse oocytes were used as the model for mechanical characterization and quality assessment. As the microgripper applies increasing loading force, the oocyte morphology progressively changes from an approximately spherical to an ellipsoidal shape (**Figure 3az**). The variation in the minor axis length of the fitted ellipse is extracted as the oocyte deformation.

The oocyte compressive deformation $\delta$ is obtained by the AI-based computer vision algorithm, which performs real-time segmentation of the oocyte contour and outputs oocyte diameter $D(t)$. The deformation is calculated as:

$$\delta = D_0 - D(t) \tag{5}$$

where $D_0$ is the oocyte diameter in the unloaded state.

To accurately characterize the mechanical behavior of oocytes over a range from small to large deformation, we adopt a stage-wise mechanical modeling strategy. Under the small-strain assumption ($\varepsilon \leq 10\%$) and assuming that the oocyte is incompressible (Poisson's ratio $v = 0.5$), the oocyte is modeled as an isotropic elastic sphere. Its force–deformation relationship follows Hertzian contact theory[27]:

$$F_{\text{cell}} = \frac{4}{3} \cdot \frac{E}{1-v^2} \cdot D_{\text{cell}}^{1/2} \cdot \delta^{3/2} \tag{6}$$

where $E$ is the effective compressive modulus, and $D_{\text{cell}}$ is the equivalent diameter of the oocyte. By performing a linear fit of the experimental data $F_{\text{cell}}$ versus $\delta^{3/2}$, the slope can be used to calculate the effective compressive modulus:

$$E = \frac{3}{4} \cdot (1-v^2) \cdot \frac{F_{\text{cell}}}{D_{\text{cell}}^{1/2} \cdot \delta^{3/2}} \tag{7}$$

Here, $E_{\text{Hertz}}$ denotes the compressive modulus obtained by nonlinear least-squares fitting, and $v$ is the Poisson's ratio. This modulus $E_{\text{Hertz}}$ represents the average stiffness of the oocyte in the quasi-linear elastic region. For larger deformations ($10\% \leq \varepsilon \leq 50\%$), we adopt an extended Tatara model based on the Mooney–Rivlin hyperelastic constitutive relationship. [28] This model accounts for both geometric nonlinearity due to finite deformation and oocyte hyperelasticity, and its core lies in describing the evolution of the compressive modulus $E$ as a function of compressive strain $\varepsilon_z$:

$$E(\varepsilon_z) = E_0 \cdot \frac{1-\varepsilon_z + \frac{\varepsilon_z^2}{3}}{(1-\varepsilon_z)^2} \tag{8}$$

where $E_0$ is compressive modulus extrapolated to zero strain. To ensure physical continuity of model parameters and fully utilize the high reliability of the small-strain data, we assign the modulus determined by the Hertz model in the $\varepsilon_z \leq 10\%$ range directly as the zero-strain modulus of the extended Tatara model, i.e.,

$$E_0 = E_{\text{Hertz}} \tag{9}$$

The theoretical basis for this assignment is that, as the strain approaches zero, the instantaneous modulus defined by the extended Tatara model, $\lim_{\varepsilon_z \to 0} E(\varepsilon_z)$, equals $E_0$ mathematically, and in the small-strain regime, the mechanical behavior of the oocyte is highly consistent with the linear elasticity assumption of the Hertz model. Therefore, setting $E_0 = E_{\text{Hertz}}$ is a reasonable and robust approximation.

Based on this mechanical model, we obtain the force response of oocytes over the full deformation range of 0–30% strain[29]. In the small-strain region, the modulus is characterized by $E_{\text{Hertz}}$, and in the large-strain region, the modulus $E(\varepsilon_z)$ is computed by substituting $E_0$ into the above equation. This method seamlessly bridges the two theoretical models, enabling an efficient and continuous description of the oocyte's mechanical behavior from linear to nonlinear regimes.

To quantify the deformation response, the AI-based computer vision algorithm automatically performs segmentation and contour extraction on each frame, obtaining the position and size of the oocyte in pixel units at every time point (Figure 3b). The results show that oocyte diameter decreases when the current is increased and recovers when the current is reduced, indicating that the cell undergoes reversible compressive deformation. When the current decreases from 18.00 mA to 15.25 mA, the cell diameter is reduced by approximately 35 μm, corresponding to a relative compressive strain of about 30% (Figures 3c and 3d).

Figure 3e summarizes the scatter distribution of forces experienced by mouse oocytes at different compressive deformations. For an individual oocyte, an increase in deformation is accompanied by a corresponding increase in $F_{\text{cell}}$. In particular, within the small-deformation regime (deformation < 10 μm), $F_{\text{cell}}$ exhibits an approximately linear increase with deformation. Under our loading configuration, the zona pellucida, cytoplasm, and other components of the oocyte are compressed simultaneously, rather than measuring only local regions. This method, therefore, ensures that the measured response reflects the compressive modulus of the entire oocyte. The maximum deformation can reach 20 μm, corresponding to compressive forces on the order of μN. Each oocyte was measured three times, and the resulting $F_{cell}$ values showed excellent reproducibility, indicating high stability of the compressive modulus measurement. The variations in the slopes of the force–deformation curves show that different oocytes exhibit distinct mechanical responses.

Building on the force-response analysis, we quantified the compressive modulus of 10 oocytes collected from different mice based on the force-deformation curve (Figure 3f). The measured moduli showed some variation across samples but generally remained within the hundreds-of-kilopascals range. Because all oocytes were collected from healthy, sexually mature female mice under normal feeding conditions, these statistics can be regarded as a reference range for normal oocyte mechanical properties. By combining data from all batches, we found that the compressive modulus of oocytes from different mice primarily falls within the range of approximately 420–800 kPa, which we define as the "normal" compressive modulus range. Oocytes with moduli significantly higher than this range are classified as "overly stiff", whereas those significantly lower are considered "overly soft". Among the 10 sample groups, most samples fall within the "normal" band (70%), while only a very small number of oocytes lie predominantly in the "overly stiff" (10%) or "overly soft" region (20%). This modulus-based classification could provide a viable strategy for assessing oocyte mechanotype at the whole-cell level.

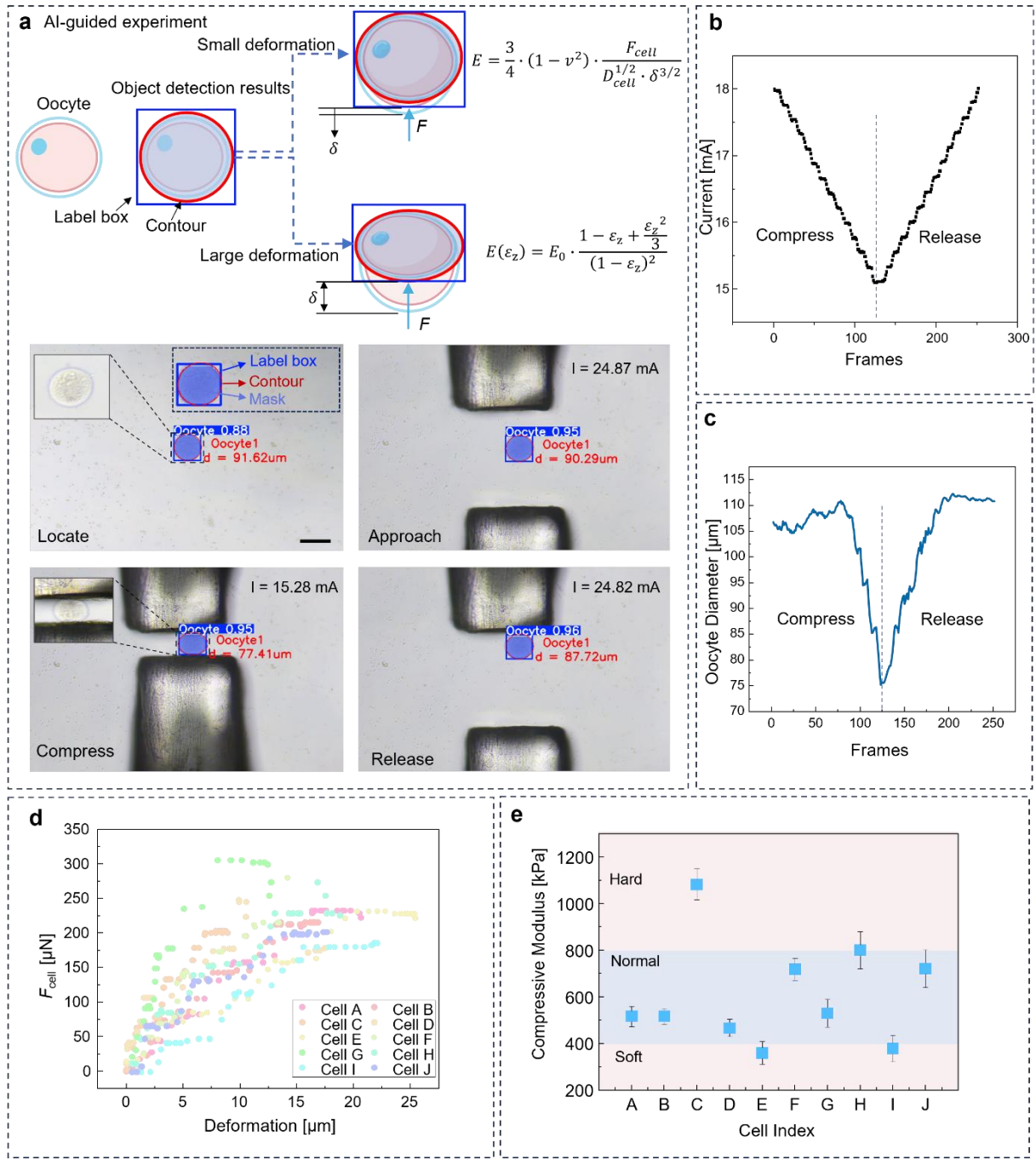

**Figure 3. Oocyte Mechanical Measurement and Quality Assessment.** a) Oocyte compressive deformation and object detection results performed on the automated microgripper. Top panel: schematic illustration; lower panel: experimental picture (scale bar = 100 μm). b) Example of electromagnet drive current over time frame during oocyte mechanical measurements. c) Example of oocyte diameter over time frame during oocyte mechanical measurements. d) Force–deformation scatter plot of mouse oocytes. e) Statistical analysis of the compressive modulus of oocytes.

**Conclusion**

We have presented an AI-guided automated oocyte quality assessment system integrated with a magnetically actuated microgripper, achieving high integration and full automation across imaging, actuation, user interaction, and mechanical measurement. The microgripper applies a defined loading force and induces micron-scale compressive deformation on oocytes enabling direct measurement of compressive modulus for oocytes. The integration of an AI-based computer vision algorithm enables automated oocyte localization, force loading, deformation extraction, and µN-level mechanical measurement, thereby substantially improving throughput and repeatability. This automated approach minimizes operator-dependent variations, ensuring high stability during repetitive operation. In addition, the modular design of the instrument facilitates rapid deployment in the IVF laboratory setting. Compared with traditional methods such as AFM and MPA, the proposed system accommodates a wider range of oocyte diameters, achieves higher automation, and enables non-destructive whole-cell mechanical assessment.

Despite its strong performance, the current system still has limitations, particularly its limited capability for evaluating small-sized cells and its restriction to single-parameter assessment. The microgripper geometry was optimized specifically for oocyte-scale cells and therefore requires redesign to accommodate smaller cell types. High-resolution 3D fabrication technology (e.g., two-photon lithography) can be used to miniaturize the microgripper and the associated electromagnetic unit in order to broaden the system's applicability to embryonic stem cells, cumulus cells, and other cells with smaller physical dimensions.[30] Furthermore, the current platform measures only mechanical deformation, whereas cellular electrical properties—such as impedance spectra—also contain valuable structural and physiological information.[31,32] To address this, integrating an impedance measurement module enables more comprehensive evaluation of oocyte developmental potential. In addition, the system currently automates only a single measurement workflow. As a next step, we aim to develop a fully autonomous intelligent experimental platform by incorporating an AI decision-making agent capable of real-time experimental planning and algorithm selection, as well as adding a robotic-arm module for automated oocyte collection. This would ultimately enable a closed-loop IVF system covering the full pipeline—from oocyte identification, force loading, and mechanical measurement to quality assessment. Such a system has the potential to further reduce human labor and advance reproductive medicine toward embodied intelligence and self-evolving experimental system.

**Experimental Section**

*Fabrication of the Microgripper*

All microgripper components were designed using SolidWorks for three-dimensional modeling. The cantilever structures and the electromagnet mounts were fabricated from PLA using a fused-filament 3D printer (Creality). The microgripper tips were produced via stereolithography (SLA) printing with a photosensitive resin (Formlabs Clear Resin). After printing, the components were rinsed with isopropanol and subsequently UV-cured for 30 min to remove residual uncured resin and improve mechanical integrity. The microgripper was then immersed in 75% ethanol for 30 min for preliminary sterilization, followed by UV irradiation for an additional 30 min to evaporate remaining solvents.

*Assembly of the Magnetic Actuation Module*

The magnetic actuation module consisted of an electromagnet, a cylindrical magnet, and a data acquisition and control unit. The electromagnet was fixed onto a custom PLA mount fabricated via 3D printing and aligned with the main optical platform to ensure a stable and uniform magnetic-field direction. Its positive and negative terminals were connected to the analog output ports of the data acquisition and control unit, enabling drive current through computer-generated analog signals. A NdFeB cylindrical magnet (diameter ≈ 2 mm) was fixed to the free end of the cantilever to ensure stable mechanical coupling and reproducible magnetic response. The magnet was oriented parallel to the external field to maximize the coupling between its magnetic moment and the applied field. A current meter with a resolution of 0.01 mA was connected in series to monitor the drive current.

*Microscopic Imaging and Actuators*

The microscopic imaging module comprises an inverted microscope (Nexcope NIB-900) with motorized objective lens and a high-resolution industrial camera (1824×1214 pixels), enabling real-time acquisition of oocyte deformation under force loading. Using microscope calibration, we obtained that 1 pixel in the image correspond to a real distance of 1.33 μm. The microscope camera supports autoexposure and frame-rate adjustment (0-50 fps). The actuation module for oocyte localization incorporates an XY-axis motorized stage with a repeatability of ±2 μm. For vertical positioning of the microgripper, a Z-axis motorized stage provides a repeatability of ±1.5 μm with a travel range of 15 mm, enabling precise Z-axis displacement control of the microgripper.

*Training of AI-based Computer Vision Algorithm*

The AI-based computer vision algorithm was built by employing YOLOv5-seg on the real-time video stream to achieve automated object detection and segmentation. The input resolution of the model is 640 × 640 pixels. The training dataset contains manually annotated oocyte images and masks, with a total of 125 images, of which 100 images were used for training (100 epochs) and the remaining 25 images for validation. Using the pretrained weights yolov5s-seg.pt, we perform transfer learning, and the model converges approximately after 100 epochs. Quantitative evaluation indicates that the model achieves an mAP@0.5 of 99.5% and a mAP@[0.5:0.95] of 89.9% for object detection, while the mask mAP@0.5 and mask mAP@[0.5:0.95] for instance segmentation reach 99.5% and 84.5%, respectively. The real-time inference speed of the model is approximately 22 fps on an RTX A5000 GPU.

*Autofocus Algorithm*

The autofocus algorithm adopts the Tenengrad gradient method to evaluate image sharpness. The Sobel horizontal and vertical gradients $G_x$ and $G_y$ are first computed, and the Tenengrad sharpness metric $T$ is defined as:

$$T = \frac{1}{N} \sum_{x,y} (G_x^2(x,y) + G_y^2(x,y)) \tag{10}$$

The algorithm uses a gradient-based strategy, in which the motorized objective is driven to adjust the focus in discrete steps (step size ≈ 10 μm), and the sharpness metric is calculated in real time. The system scans over four steps above and below the current focal plane and selects the plane with the maximum sharpness as the working focal plane.

*Voice Interaction Module*

The voice interaction module used a microphone and Google Gemini-2.5 Pro application programming interface (API) for natural language recognition. The model was first provided with a detailed description of the experimental background, including the experiment name, instrument setup, and the set of possible voice commands. Based on this prior context, it performs fuzzy matching of spoken instructions and identifies task-related keywords, which can be mapped to a predefined workflow for oocyte mechanical measurement.

*Collection and Preparation of Oocytes*

The oocytes used in this study were obtained from female mice aged 8–11 weeks. After superovulation was induced with pregnant mare serum gonadotropin (PMSG), the mice were sacrificed, and oocytes were directly collected by oviduct dissection. The oocytes were then immersed in pre-equilibrated Human Tubal Fluid (HTF) medium for approximately 15 minutes to facilitate their natural detachment from the follicular tissue. Subsequently, under microscopic observation, the released oocytes were gently aspirated using a micropipette, and cellular debris and residual follicular material were carefully removed to ensure that only morphologically intact oocytes with zona pellucida were retained. The collected oocytes were then transferred into a culture dish containing tubal fluid to maintain their physiological state. The entire procedure was carried out on the motorized microscope stage.

**Acknowledgements**

H.X. conceived the idea and designed the experiments. Y.G., W.Z. and X.S. performed the experiments with the help of J.H., X.C. and X.L. Y.G., W.Z., and X.S. analyzed the experimental data with the help of H.X. and Y.L. All authors discussed the results and contributed to the writing of the manuscript. Y.G., W.Z. and X.S. contributed equally to this work.

**Data Availability Statement**

Source data are provided with the journal version of this paper. Other data are available from the corresponding authors upon reasonable request.